\documentclass[conference]{IEEEtran}
\IEEEoverridecommandlockouts

\usepackage{comment}

\usepackage{graphicx}
\usepackage{cite}
\usepackage{amsmath,amssymb,amsfonts}
\usepackage{algorithmicx}
\usepackage{textcomp}
\usepackage{xcolor}

\usepackage{amsthm}

\usepackage[caption=false]{subfig}

\usepackage{authblk}

\usepackage{algorithm,algpseudocode}

\usepackage{multicol}

\usepackage{mathtools}

\usepackage{cuted}

\usepackage{color,soul}
\def\BibTeX{{\rm B\kern-.05em{\sc i\kern-.025em b}\kern-.08em
    T\kern-.1667em\lower.7ex\hbox{E}\kern-.125emX}}

\makeatletter
\newcommand{\linebreakand}{%
\end{@IEEEauthorhalign}
\hfill\mbox{}\par
\mbox{}\hfill\begin{@IEEEauthorhalign}
}

\makeatother

\begin{document}

\title{
Dynamic Power Allocation in OFDM ISAC for Time of Arrival Estimation

\thanks{This work has been supported by the Smart Networks and Services Joint Undetaking (SNS JU) project 6G-DISAC under the EU’s Horizon Europe research and innovation program under Grant Agreement no. 101139130}
}

\author[1]{Ali Al Khansa}
\author[2]{Giyyarpuram Madhusudan}
\author[2]{Guillaume Larue}
\author[2]{Louis-Adrien Dufrène}

\affil[1]{Orange Labs, Rennes, France}
\affil[2]{Orange Labs, Meylan, France}

\affil[ ]{Emails: \{ali.alkhansa;giyyarpuram.madhusudan;guillaume.larue;louisadrien.dufrene\}@orange.com}

\maketitle

\IEEEpubidadjcol

\begin{abstract}

Resource allocation in Integrated Sensing and Communication (ISAC) systems is critical for balancing communication and sensing performance. This paper introduces a novel dynamic power allocation strategy for Orthogonal Frequency Division Multiplexing (OFDM) ISAC systems, optimizing communication capacity while adhering to Peak Side-lobe Level (PSL) and sensing accuracy constraints, particularly for Time of Arrival (ToA) estimation. Unlike conventional methods that address either PSL or accuracy in isolation, our approach dynamically allocates power to satisfy both constraints. Additionally, it prioritizes communication when sensing performance is insufficient, avoiding any loss in communication capacity. Numerical results validate the importance of considering both sensing constraints and demonstrate the effectiveness of the proposed dynamic power allocation strategy.

\end{abstract}

\vspace{0.5cm}

\section{Introduction}

\IEEEPARstart{I}{\lowercase{n}}tegrated Sensing and Communication (ISAC) is an emerging paradigm in wireless communication that combines communication and sensing in a unified system. The primary motivation behind ISAC is to enhance spectral and energy efficiency by allowing a single waveform to serve dual purposes. This integration is especially pertinent as we advance toward the sixth Generation (6G) of wireless networks \cite{liu2022integrated}, where the demand for both high-speed communication and precise sensing capabilities is expected to significantly increase. One of the key sensing parameters we focus on is the estimation of the Time of Arrival (ToA), which is crucial for effective distance measurement, and significantly influences both the design and optimization of ISAC systems.

Resource allocation, particularly power allocation over Orthogonal Frequency Division Multiplexing (OFDM) frames (e.g., \cite{keskin2021peak,driusso2014performance,montalban2013power}), is a critical aspect of ISAC system design. OFDM is the current framework used in 5G networks and is anticipated to remain a cornerstone in 6G networks. Accordingly, it has been widely applied in the design of communication centric ISAC waveforms \cite{wei20225g}. Efficient resource allocation is essential to balance the dual objectives of ISAC systems, ensuring that both communication and sensing performances are optimized. In the context of ISAC, this involves navigating the trade-offs between the performance of the communication (e.g., capacity) and the sensing (e.g., accuracy). Trade-offs in ISAC systems often arise because the requirements for optimal communication performance can conflict with those for optimal sensing performance. For instance, maximizing communication capacity might degrade sensing accuracy and vice versa. Effective power allocation strategies must therefore account for these trade-offs to ensure that neither communication nor sensing capabilities are compromised, thereby achieving a balanced and synergistic optimization.

From the State-of-the-Art (SotA) of ISAC systems, two main categories can be seen \cite{liu2022integrated}: Sensing-Centric Design
(SCD) \cite{ahmed2023sensing}, and Communication-Centric Design (CCD) \cite{barneto2019full}. The SCD focuses on embedding communication data into radar signals while maintaining the integrity of the sensing function. This approach emphasizes the sensing parameters and primarily assesses the data-carrying capacity of the signal without detailed optimization of the communication channel. The CCD on the other hand aims to maximize the information content derived from radar signals after they reflect off targets. This approach optimizes communication capacity while imposing constraints on the sensing accuracy (mainly using the lower bound: Cramér-Rao Bound (CRB) as in \cite{montalban2013power}). However, it often overlooks ambiguity analysis and Peak Side-lobe Level (PSL) minimization. The PSL of the autocorrelation function is an important performance criterion that is crucial for avoiding false positives and preventing the sidelobes of strong targets from masking weaker targets \cite{keskin2021peak}.

In \cite{wymeersch2022radio}, the authors highlight the trade-off between ambiguity and accuracy in sensing. They illustrate how different power allocations in an OFDM ISAC system can lead to significant variations in sensing metrics. This aspect is often overlooked in many SotA methods, which primarily focus on accuracy, often quantified by the CRB, as the sole metric for sensing performance. A similar observation was made in \cite{driusso2014performance}, where the authors noted that, intuitively, to increase the lower bound on accuracy, power should be concentrated at the edges of the available bandwidth. However, they emphasize that this bound is only tight in the asymptotic regime. 

To our interest, the work of \cite{keskin2021peak} introduces a joint sensing and communication waveform design method that optimizes power allocation over OFDM carriers with constraints on PSL. It bridges the gap between the SCD and the CCD approaches by considering both radar target detection in a multi-target setting and capacity optimization. However, in this paper, we argue that while considering the ambiguity analysis and the PSL is important, it is not sufficient on its own as it does not guarantee overall sensing accuracy. Additionally, formulating the resource allocation problem solely based on maximizing capacity under a PSL constraint can result in suboptimal communication performance. The capacity degradation resulting from meeting this constraint could be substantial when compared to the optimal capacity without sensing considerations. This significant trade-off might render the solution unacceptable in practice.

Therefore, we propose a novel dynamic power allocation strategy that aims to prevent significant losses in communication performance due to impractical sensing parameters. Specifically, in an OFDM ISAC system, we optimize the communication capacity while imposing constraints on both ambiguity (e.g., PSL) and sensing accuracy, as well as constraints on capacity loss due to sensing activities. The novelty of this approach lies in its ability to dynamically maintain optimal communication capacity when any of the sensing constraints are not met or when the capacity cost to satisfy the constraints is too high, thereby avoiding unnecessary degradation in communication performance. To achieve this, the proposed algorithm begins with an optimization for communication capacity, subsequently adjusting the power allocation dynamically to satisfy both communication and sensing requirements as per the defined constraints. Numerical simulations are presented to validate the effectiveness of the proposed strategy.

The rest of the paper is organized as follows: Section II details the specific parameters and signal structures of the OFDM ISAC system. Section III outlines the formulation of the optimization problem. Section IV describes the algorithmic steps of the dynamic power allocation process. Section V analyzes the simulation outcomes, illustrating the practical effectiveness of our proposed dynamic algorithm. Finally, Section VI concludes the work.

\section{OFDM ISAC Parameters}
This paper examines how allocating power differently impacts both the sensing parameters (accuracy and ambiguity) and the communication capacity. To do so, we focus on the capacity and the estimation of the ToA (which is equivalent to distance estimation). Note that other estimation parameters can be considered for the sensing part with no loss of generality of the main proposal of a dynamic power allocation which decides to do either ISAC or communication only.

\subsection{OFDM ISAC Signal Structure}

We consider a continuous OFDM transmission scheme where symbols include cyclic prefixes. We assume the presence of a Line-Of-Sight (LOS) propagation path alongside other propagation paths involving reflections from scattering surfaces in the environment. Additionally, we assume that the communication channel experiences slow fading and remains constant over several OFDM symbols. On the receiving end, the receiver is configured to align its reception windows precisely with the incoming OFDM symbols. Radar receivers rely on a technique called a matched filter. This filter maximizes the signal's strength compared to white noise for optimal detection. In the context of OFDM radar, OFDM pulse compression is achieved by applying a matched filter in the frequency domain \cite{tigrek2012ofdm}. For this specific OFDM pulse compression method, we begin by considering an OFDM radar signal that bounces back from a single target at time $t$ \cite{tigrek2012ofdm}:

\begin{equation}
    s(t) = \sum_{k = 0}^{K-1} \alpha \sqrt{P_{k}}x_{k} e^{ j 2 \pi k \Delta f ( t - \tau ) },\textcolor{white}{aaaa} 0 \le t < T,
\end{equation}
The parameter $\alpha$ represents the complex target coefficient, including path loss and radar cross-section. The index $k$ refers to the sub-carrier index, ranging from 0 to $K-1$, where $K$ is the total number of sub-carriers. The term $P_{k}$ denotes the power allocated to the $k^{\text{th}}$ sub-carrier, and $x_{k}$ represents the symbol transmitted on that sub-carrier. The delay of the radar echo is denoted by $\tau$, and the OFDM carrier spacing is given by $f=1/T$ where $T$ represents the symbol duration. The formulation assumes that the cyclic prefix duration for the OFDM symbol is greater than the delay $\tau$.

The OFDM pulse compression operates on the sampled echo by applying a Discrete Fourier transform (DFT), multiplying each carrier with the complex conjugate of its communication payload and applying an inverse discrete Fourier transform (IDFT) to obtain the pulse compression output, which can be formulated as:

\begin{equation} 
    r_{n} = \sum_{k = 0}^{K-1} P_{k} |x_{k}|^{2} e^{ -j 2 \pi k \Delta f \tau } e^{ j 2 \pi \frac{nk}{N} } + w_{n},\textcolor{white}{aa} 0 \le n \le N-1,
\label{CompOut}
\end{equation}
where $N \ge K$ represents the number of samples and $\alpha$ is omitted as it only scales the entire pulse compression output, and $w_{n}$ refers to the Gaussian noise. The Doppler effect is not considered in the two equations above. The range sidelobes for non-zero Doppler would depend on both the power allocation and the phases of the OFDM sub-carriers \cite{keskin2021peak}. However, this study focuses on a scenario where the Doppler effect has minimal influence on the sub-carrier phases (e.g., stationary or slow-moving targets), allowing us to simplify the analysis. In this case, the delay profile obtained from the OFDM waveform is the inverse Fourier transform of the power spectrum, which is governed by the amplitudes of the OFDM sub-carriers. Although the complex exponential term in the second equation shifts the delay profile by $\tau$, it does not significantly affect the range sidelobes.


\subsection{Communication Metric: Capacity}
Given complex channel gains $\textbf{h}=[h_{0},...,h_{K-1}]^{T}$ across the sub-carriers, and assuming that the fading channels are well known, the channel capacity is calculated through the Shannon-Hartley theorem:

\begin{equation}\label{capacity}
    C(\textbf{p;h}) = \sum_{k=0}^{K-1} \log_{2} \left\{ 1+\frac{P_{k}|h_{k}|^{2}}{N_{0}\Delta f } \right\},
\end{equation}
where:
\begin{itemize}
    \item $N_{0}$ is the noise power spectral density.
    \item $\textbf{p} = [P_{0},...,P_{K-1}]^{T}$  is the power allocation.
\end{itemize}
When the transmitter has knowledge of the channel gains, the capacity can be optimized with respect to \textbf{p}, resulting in the well-known Water-Filling (WF) solution \cite{cover1999elements}.

\subsection{Sensing Metrics: PSL and Accuracy}
As highlighted in \cite{liu2022integrated}, unlike communication systems where capacity serves as a clear metric for performance evaluation, sensing systems lack a directly analogous ``capacity" metric. This absence underscores the complexity of defining and measuring the efficiency and effectiveness of sensing systems, which depend on multiple, often conflicting performance metrics. This complexity guides our focus on two critical aspects of sensing in ISAC systems: the PSL and the accuracy, which serve as proxy metrics for assessing the reliability and accuracy of sensing operations.
\subsubsection{PSL}
The PSL refers to the largest amplitude of the side lobes in the radar signal's autocorrelation function (or its matched filter response), relative to the main lobe (check Fig. \ref{fig:NRfig_0b} in the numerical results section for an example). Strong side lobes increase the ambiguity and can lead to false positives, and thus, we aim to have some upper bound constraints on the PSL. The pulse compression output in (\ref{CompOut}) can be represented as
\begin{equation}\label{r_compact}
\mathbf{r} = \mathbf{W}^H (\mathbf{p} \odot \mathbf{x})
\end{equation}
where $\odot$ represents the Hadamard product, and $\mathbf{r} = [r_0, \ldots, r_{N-1}]^T$ corresponds to the pulse compression output. The matrix $\mathbf{W} \in \mathbb{C}^{K \times N}$ has elements defined as $w_{k,n} = e^{-j2\pi k \left( \frac{n}{N} - \Delta f \tau \right)}$, while $\mathbf{x} = [|x_0|^2, \ldots, |x_{K-1}|^2]$ refers to the transmitted symbols. The PSL of the pulse compression output $\mathbf{r}$ can be defined as
\begin{equation}
\text{PSL} = \max_{n \in \mathcal{S}} |r_n|^2
\end{equation}
where $\mathcal{S} \subseteq \{0, \ldots, N-1\}$ denotes the sidelobe region of interest. This region may encompass all sidelobes or specific subsets around the main lobe. As mentioned in \cite{wymeersch2022radio}, by leveraging side information and focusing solely on the specified regions, we can disregard ambiguities outside the region of interest. However, in this study, without loss of generality, we assume no side information is available, and thus all sidelobes are considered.

Equation (\ref{r_compact}) shows that the PSL of the pulse compression output is influenced by the power allocation $\mathbf{p}$ and the magnitude of the random data $\mathbf{x}$. Given that radar processing takes place across several OFDM symbols while the channel remains constant, the PSL of the average pulse compression output becomes an important metric:
\begin{equation}
\max_{n \in \mathcal{S}} \left| \mathbb{E}(r_n) \right|^2
\end{equation}
where the expectation is over the distribution of $\mathbf{x}$. We assume independent and identically distributed, zero-mean unit-variance complex Gaussian data symbols, i.e., $x_k \sim \mathcal{CN}(0,1)$. The expected pulse compression output is then given by:
\begin{equation}
\label{expPulse}
\mathbb{E}(\mathbf{r}) = \frac{1}{2} \mathbf{W}^H \mathbf{p}.
\end{equation}

\subsubsection{Accuracy}

Concerning the accuracy, different metrics can be considered, but the most famous one is the CRB. The CRB provides a theoretical lower bound on the variance of any unbiased estimator. The CRB quantifies the best possible precision for parameter estimation in noisy conditions: a lower CRB indicates the potential for more accurate estimates.

It is well known that to minimize the CRB for a given bandwidth, it is optimal to put more power on the edges of the available sub-carriers \cite{driusso2014performance}. Thus, in order to improve the CRB, we would be trying to maximize the following:
\begin{equation}
\sum_{k=0}^{K-1} \left( k - \frac{K}{2} \right)^2 P_k.
\end{equation}
Another accuracy consideration would be simply checking the width of the main lobe and which we will mention more in the numerical result section.

In practice, accuracy can be computed through Monte Carlo simulations and evaluated using metrics like mean square error. Moreover, accuracy depends not only on the power allocation but also on the strategy employed for estimation (e.g., periodogram-based algorithms \cite{braun2014ofdm}). With this in mind, we emphasize that the focus of this paper is on the effect of power allocation on sensing and communication parameters, irrespective of the accuracy metric used (e.g., CRB or Main Lobe Width (MLW)) or the estimation algorithm applied. Similarly, for ambiguity, we concentrate on the PSL, as it is heavily influenced by power allocation, though our dynamic algorithm can be adapted to accommodate other ambiguity metrics.

\section{Problem Formulation}

In ISAC systems, much of the prior research has focused on optimizing accuracy, often using the CRB as a key metric (e.g., \cite{montalban2013power}). However, relying solely on that is insufficient, as it overlooks the importance of ambiguity analysis for distinguishing strong and weak target signals, especially at low Signal to Noise Ratio (SNR). Therefore, \cite{keskin2021peak} proposes a shift towards optimizing PSL to ensure robust sensing.

Building on this, our objective is twofold: first, to incorporate both accuracy and ambiguity constraints in ISAC resource allocation, as optimizing just one is inadequate. Second, we aim to develop a strategy that balances communication capacity with sensing parameters, as focusing solely on capacity may yield impractical results. To address these issues, we propose a dynamic power allocation strategy that maximizes communication capacity while considering accuracy (e.g., CRB), ambiguity (e.g., PSL), and allowable capacity loss compared to the WF algorithm. This is formalized as:
\begin{equation}
\begin{aligned}
    & \max_{\mathbf{p}} \quad C(\mathbf{p}; \mathbf{h}) = \sum_{k=0}^{K-1} \log_2 \left( 1 + \frac{P_k |h_k|^2}{N_0 \Delta f} \right) \\
    & \text{s.t.} \quad \sum_{k=0}^{K-1} \left( k - \frac{K}{2} \right)^2 P_k \geq \gamma_{\text{CRB}} \\
    & \quad \; \max_{n \in S} |\mathbf{w}_n^H \mathbf{p}|^2 \leq \gamma_{\text{PSL}} \\
    & \quad \; \frac{C_{\text{WF}}-C}{C_{\text{WF}}} \leq \gamma_c \\
    & \quad \; \mathbf{1}^T \mathbf{p} = P_T, \quad \mathbf{p} \geq 0
\end{aligned}
\end{equation}

In this formulation, the first constraint ensures accuracy (CRB), the second addresses ambiguity (PSL), the third limits capacity loss, and the last enforces the power budget $P_{T}$ and non-negative allocated powers. The values of $\gamma$ in the first three constraints are selected based on the considered use case.

\section{Proposed Approach}

As the problem is no longer convex and can't be solved using WF, we propose next some heuristic alternatives. A straightforward algorithm is the following:
\begin{enumerate}
    \item Compute the optimal capacity \(C_{\text{WF}}\) using the WF strategy without considering any sensing parameters.
    \item Calculate the suboptimal capacity \(C_{\text{PSL}}\) while incorporating PSL constraints as outlined in \cite{keskin2021peak}. Naturally, \(C_{\text{PSL}}\) is less than or equal to \(C_{\text{WF}}\):
    $$C_{\text{PSL}} \leq C_{\text{WF}}.$$
    \item Calculate the capacity loss ratio \(C_{\text{loss}} = \frac{C_{\text{WF}}-C_{\text{PSL}}}{C_{\text{WF}}}\) and determine the CRB.
    \item If \(C_{\text{loss}}\) and CRB are below some certain thresholds (i.e., \(C_{\text{loss}} < \gamma_c\), CRB \(< \gamma_{\text{CRB}}\)):
    \begin{itemize}
        \item Use power allocation as described in \cite{keskin2021peak}, i.e., perform ISAC.
    \end{itemize}
    \item Otherwise, stick to the WF strategy, allocating power to optimize capacity only, resulting in \(C_{\text{WF}}\), i.e., perform communication only.
\end{enumerate}
The strategy focuses on reducing capacity loss, satisfying PSL constraints, and maintaining accuracy via CRB verification.

\subsection*{Proposed Algorithm}

\begin{algorithm}[t]
\caption{The Dynamic Power Allocation Algorithm.}\label{alg}
\begin{algorithmic}[1]
\State According to $\mathbf{h}$, compute $\textbf{p}_{\text{WF}}$ following the WF algo 
\State Compute the sensing parameters: PSL and CRB 
\If{($\text{PSL} > \gamma_{\text{PSL}}$ and $\text{CRB} > \gamma_{\text{CRB}}$)} 
    \State $\textbf{p} = \textbf{p}_{\text{WF}}$ \Comment{No ISAC}
\Else
    \If{$\text{PSL} < \gamma_{\text{PSL}}$}                                             \Comment{If CRB is not satisfied}    
        \State $\textbf{p} = \textit{Binary\_Search\_CRB}(\textbf{p}_{\text{WF}},\textbf{p}_{\text{edges}})$       
    \Else                                                                               \Comment{If PSL is not satisfied}    
        \State $\textbf{p} = \textit{Binary\_Search\_PSL}(\textbf{p}_{\text{WF}},\textbf{p}_{\text{PSL\_Opt}})$     
    \EndIf
\EndIf
\end{algorithmic}
\end{algorithm}

Due to the limitations of the initial approach, we propose an enhanced strategy that integrates both communication and sensing parameter constraints. This approach is still dynamic and communication-centric. We begin with the WF algorithm and then modify the allocation by considering sensing constraints. The enhanced approach leverages three different allocations: WF allocation (optimal for capacity), PSL optimal allocation (the allocation when minimizing the PSL with no CRB or capacity constraints as in \cite{keskin2021peak}), and CRB optimal allocation (edges-only approach). This proposal builds on the intuition that usually, when trying to optimize the accuracy (i.e., the CRB), the ambiguity gets higher. Thus, trying to satisfy the PSL constraint will come at the cost of both the accuracy and the capacity. 

Accordingly, this approach starts with WF allocation, and then tests the two sensing parameters (i.e., the CRB and the PSL) when using the WF allocation. Then, this approach decides to do communication only if both sensing parameters are not satisfied. The reasoning is that trying to improve one of the sensing parameters usually comes at the cost of the performance of the second sensing parameter. Now, if one of the parameters is satisfied, this approach tries to satisfy the second parameter by modifying the WF allocation using binary search approach with the PSL optimal allocation (to improve PSL) or with CRB optimal (i.e., edges-only allocation) (to improve the CRB). Finally, this approach performs ISAC if it was possible to satisfy the two sensing parameters while not breaking the capacity loss constraint. The steps of this approach are summarised below and presented in Algo. \ref{alg}.

\begin{algorithm}[t] 
\caption{$\textit{Binary\_Search\_CRB}(\textbf{p}_{\text{WF}},\textbf{p}_{\text{edges}})$}\label{alg:binary_search_crb}
\begin{algorithmic}[1]
\State Fix $\epsilon$ (e.g., 0.01), $\text{left} \gets 0$, $\text{right} \gets 0$, $C_{\text{max}}\gets 0$, $\alpha_{\text{sol}}\gets 0$
\State $\alpha \gets \text{left} + (\text{right} - \text{left})/2$
\While{$(\text{left} - \text{right} > \epsilon)$}
    \State $\textbf{p} \gets \alpha \cdot \textbf{p}_{\text{edges}} + (1 - \alpha) \cdot \textbf{p}_{\text{WF}}$
    \State Compute the metrics $C$, PSL, and CRB
    \If{$(C > C_{\text{max}} \ \text{and} \ \text{PSL} < \gamma_{\text{PSL}} \ \text{and} \ C_{\text{loss}} < \gamma_{C})$}
        \State $C_{\text{max}} \gets C$, $\alpha_{\text{sol}}\gets \alpha$
        \State $\text{left} \gets \alpha$
        \State $\alpha \gets \text{left} + (\text{right} - \text{left})/2$
    \Else
        \State $\text{right} \gets \alpha$
        \State $\alpha \gets \text{left} + (\text{right} - \text{left})/2$
    \EndIf
\EndWhile
    \State \Return $\textbf{p} = \alpha_{\text{sol}} \cdot \textbf{p}_{\text{edges}} + (1 - \alpha_{\text{sol}}) \cdot \textbf{p}_{\text{WF}}$
\end{algorithmic}
\end{algorithm}

\begin{enumerate}
    \item Start with WF allocation.
    \item Check PSL and CRB constraints:
    \begin{itemize}
        \item If neither is satisfied, stick to WF with no ISAC.
        \item If PSL is satisfied but CRB is not (or even if PSL is satisfied and we would like to improve the CRB at the cost of capacity degradation), adjust WF using a binary search strategy with weighted values of the edges-only allocation while ensuring capacity and PSL constraints (Algo. \ref{alg:binary_search_crb}).
        \item If CRB is satisfied but PSL is not, adjust WF similarly using PSL optimization (similar to Algo. \ref{alg:binary_search_crb} but omitted for brevity).
    \end{itemize}
    \item Perform ISAC if both sensing parameters are satisfied without violating the capacity constraint.
\end{enumerate}
This approach dynamically decides between ISAC and communication, ensuring efficient resource utilization.

\section{Numerical Results}

In this section, we validate our proposed dynamic power allocation strategy via numerical simulations using an OFDM model with 128 sub-carriers and a bandwidth of 1 MHz. Our objective is to demonstrate the importance of integrating both PSL and accuracy constraints along with capacity loss considerations, as opposed to focusing solely on maximizing capacity under a single constraint.

\begin{figure}
    \centering
    \subfloat[Power Allocation.]{%
        \includegraphics[width=0.4\textwidth]{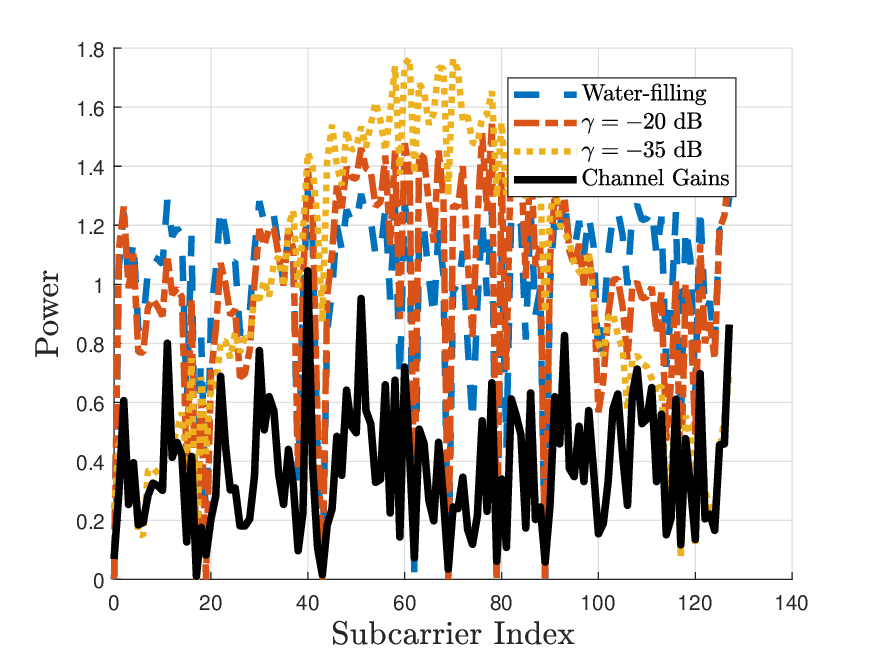}
        \label{fig:NRfig_0a}
    }
    \hfill
    \subfloat[Range Profile (IFFT Output).]{%
        \includegraphics[width=0.4\textwidth]{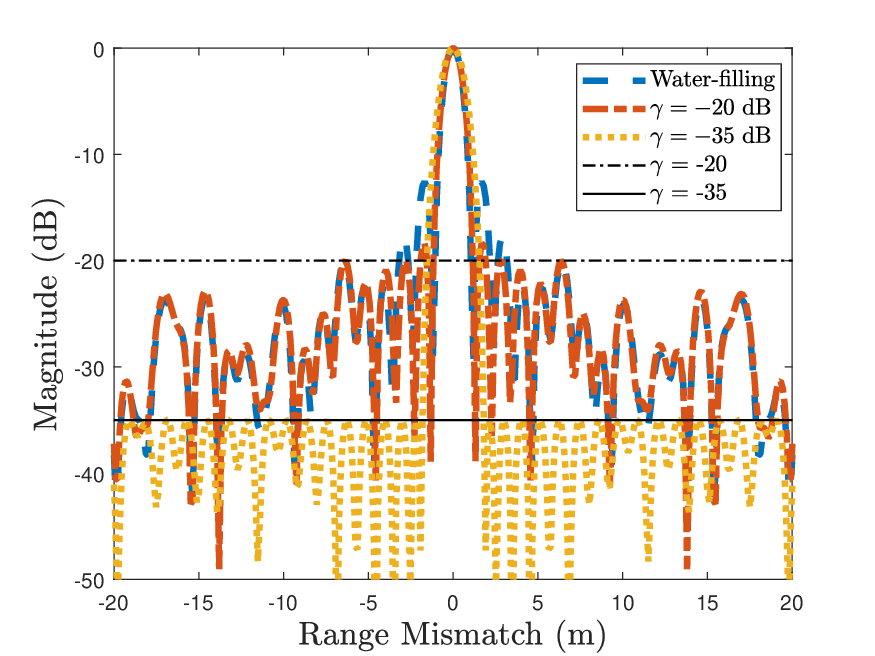}
        \label{fig:NRfig_0b}
    }
    \caption{Comparative analysis of power allocation strategies on the IFFT output for the OFDM ISAC ToA estimation with different PSL constraints.}
    \label{fig:combined_figure}
\end{figure}

To motivate our proposal, we first revisit the solution presented in \cite{keskin2021peak}, which maximizes the capacity under the PSL constraint alone. As shown in Fig. \ref{fig:combined_figure}, we display the sub-carrier powers $\mathbf{p}$ and the channel gains $|h_{k}|^{2}$ (Fig. \ref{fig:NRfig_0a}), and the corresponding expected pulse compression outputs in (\ref{expPulse}) (Fig. \ref{fig:NRfig_0b}) for various PSL constraints. Consistent with previous findings, the optimal power allocation resembles the WF solution under loose PSL thresholds. However, as the PSL threshold tightens, the power allocation assumes a window-like shape, aligning with conventional windowing techniques \cite{braun2014ofdm}.

   \begin{figure}
    \centering
    \includegraphics[scale=0.4]{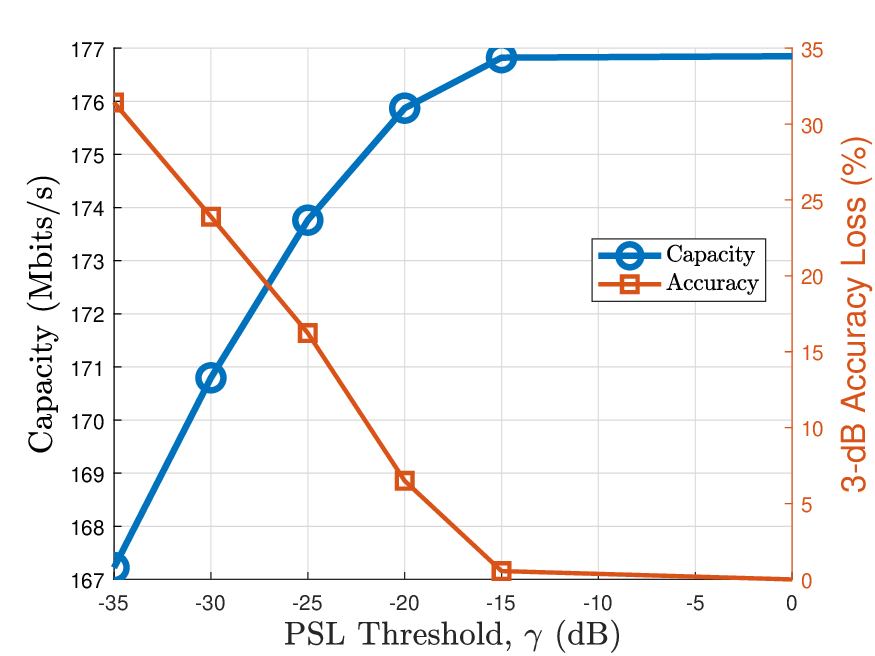}
    \caption{Trade-offs between capacity and accuracy in OFDM ISAC Systems under different PSL constraints.}
    \label{fig:NRfig_1}
    \end{figure}

In addition to capacity reduction due to the PSL constraint, an important observation is that the MLW increases with tighter PSL constraints (\ref{fig:NRfig_0b}), leading to a decrease in accuracy. Accordingly, we present two curves in Fig. \ref{fig:NRfig_1}. The first curve represents the capacity reduction, while the second illustrates the 3-dB accuracy loss, defined as the percentage increase in the MLW. This loss is calculated as the percentage increase in the MLW, defined as $(x_2 - x_1)/x_1 \times 100$ where \(x_1\) is the width of the main lobe at 3 dB without any PSL constraint, and \(x_2\) is the width with the PSL constraint applied. As depicted in Fig. 2, for loose PSL constraints, both accuracy and capacity are conserved. However, tightening the PSL constraint results in the degradation of these performances, leading to a loss of approximately 12 Mbits in capacity and around 35\% in accuracy. This trade-off between PSL, accuracy, and capacity highlights the need for a strategy that balances communication and sensing metrics without compromising one for the other. 

\begin{figure}
    \centering
    \subfloat[Accuracy Optimization: As $\alpha$ transitions from 0 to 1 (WF to accuracy-optimal), capacity (blue) worsens, suppression (red) worsens, but the MLW (yellow) improves.]{
    
        \includegraphics[width=0.45\textwidth]{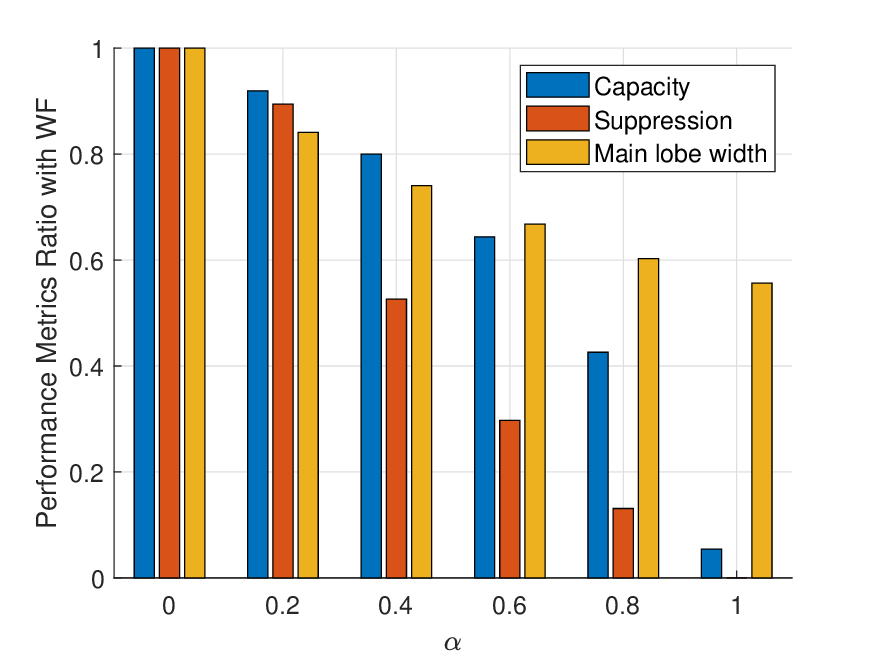}
        \label{fig:NRfig_2a}
    }
    \hfill
    \subfloat[PSL Optimization: As $\alpha$ transitions from 0 to 1 (WF to PSL-optimal), capacity (blue) worsens, the MLW (yellow) worsens, but suppression (red) improves.]{
        \includegraphics[width=0.45\textwidth]{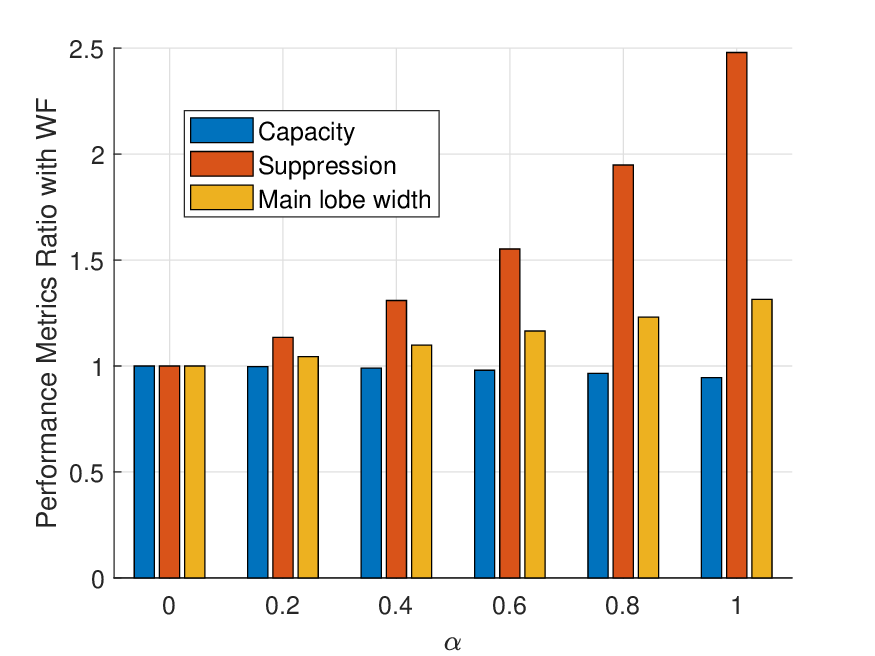}
        \label{fig:NRfig_2b}
    }
    \caption{Comparative analysis of weighting factor $\alpha$ effects on OFDM ISAC system metrics under PSL and accuracy optimization strategies.}
    \label{fig:combined_figure_2}
\end{figure}

   \begin{figure}
    \centering
    \includegraphics[scale=0.43]{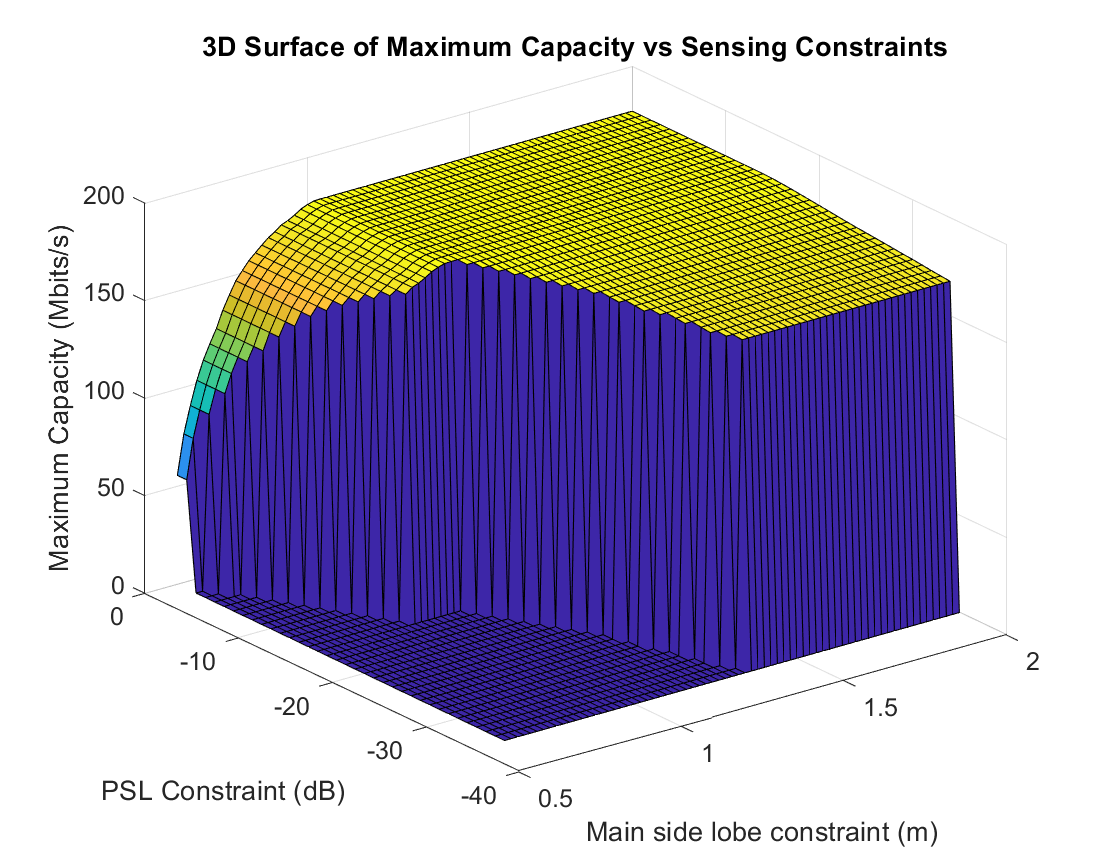}
    \caption{3D visualization of maximum capacity under varying PSL (i.e., $\gamma_{\text{PSL}}$) and accuracy constraints in OFDM ISAC System.}
    \label{fig:NRfig_3}
    \end{figure}

Our proposal tackles the challenge of balancing communication and sensing by using a binary search strategy to dynamically adjust power allocation, optimizing both communication capacity and sensing accuracy simultaneously (Algo. \ref{alg:binary_search_crb} for CRB and similarly for PSL). Fig. \ref{fig:combined_figure_2} shows histograms of the effect of the parameter $\alpha$ (the weighting factor) on key performance metrics: communication capacity, PSL, and sensing accuracy. The capacity metric compares different $\alpha$ values to the baseline water-filling method ($\alpha = 0$). The PSL metric reflects the suppression ratio, defined as the difference between the PSL and MLW, while the accuracy metric shows the ratio of the MLW for different $\alpha$ values.

In Fig. \ref{fig:NRfig_2a} (accuracy optimization), we observe that prioritizing sensing accuracy (minimizing MLW in yellow) significantly reduces PSL suppression and capacity (in red and blue), especially at $\alpha = 1$, where suppression is minimal, and capacity is low. Conversely, Fig. \ref{fig:NRfig_2b} (PSL optimization) shows that focusing on PSL results in larger MLW and reduced accuracy. So, we see that both optimizations come at a capacity cost. This highlights the trade-offs between capacity, suppression, and accuracy, emphasizing the effectiveness of our dynamic allocation method in achieving a balanced performance.


Finally, Fig. \ref{fig:NRfig_3} shows a 3D surface plot of maximum achievable capacity under varying PSL and MLW constraints. This illustrates how our dynamic power allocation strategy adapts to different sensing needs while maintaining communication performance. When PSL is relaxed (e.g., above -10 dB) and the MLW is broad (e.g., $>$1.5 m), the system achieves near-optimal capacity. However, under strict constraints—such as PSL below -30 dB or a MLW under 0.5 m—it cannot meet sensing requirements with the available bandwidth. In extreme cases, ISAC becomes challenging, forcing the system to prioritize communication or sensing. This flexibility allows our approach to dynamically adjust, ensuring efficient ISAC operation or shifting to communication-only mode when joint operation is unfeasible.

\section{Conclusions}

In conclusion, this work introduces a dynamic power allocation strategy for OFDM ISAC systems that addresses both ambiguity and accuracy constraints. Unlike traditional methods focusing on a single constraint, our approach optimizes communication capacity without sacrificing sensing performance. By addressing potential capacity loss from stringent sensing requirements, our strategy provides a balanced solution. Numerical results highlight the trade-offs and effectiveness of dynamic allocation in managing competing demands.

\bibliography{conf1}

\begin{thebibliography}{10}
\providecommand{\url}[1]{#1}
\csname url@samestyle\endcsname
\providecommand{\newblock}{\relax}
\providecommand{\bibinfo}[2]{#2}
\providecommand{\BIBentrySTDinterwordspacing}{\spaceskip=0pt\relax}
\providecommand{\BIBentryALTinterwordstretchfactor}{4}
\providecommand{\BIBentryALTinterwordspacing}{\spaceskip=\fontdimen2\font plus
\BIBentryALTinterwordstretchfactor\fontdimen3\font minus \fontdimen4\font\relax}
\providecommand{\BIBforeignlanguage}[2]{{%
\expandafter\ifx\csname l@#1\endcsname\relax
\typeout{** WARNING: IEEEtran.bst: No hyphenation pattern has been}%
\typeout{** loaded for the language `#1'. Using the pattern for}%
\typeout{** the default language instead.}%
\else
\language=\csname l@#1\endcsname
\fi
#2}}
\providecommand{\BIBdecl}{\relax}
\BIBdecl

\bibitem{liu2022integrated}
F.~Liu, Y.~Cui, C.~Masouros, J.~Xu, T.~X. Han, Y.~C. Eldar, and S.~Buzzi, ``Integrated sensing and communications: Toward dual-functional wireless networks for 6g and beyond,'' \emph{IEEE journal on selected areas in communications}, vol.~40, no.~6, pp. 1728--1767, 2022.

\bibitem{keskin2021peak}
M.~F. Keskin, R.~F. Tigrek, C.~Aydogdu, F.~Lampel, H.~Wymeersch, A.~Alvarado, and F.~M. Willems, ``Peak sidelobe level based waveform optimization for ofdm joint radar-communications,'' in \emph{2020 17th European Radar Conference (EuRAD)}.\hskip 1em plus 0.5em minus 0.4em\relax IEEE, 2021, pp. 1--4.

\bibitem{driusso2014performance}
M.~Driusso, M.~Comisso, F.~Babich, and C.~Marshall, ``Performance analysis of time of arrival estimation on ofdm signals,'' \emph{IEEE Signal Processing Letters}, vol.~22, no.~7, pp. 983--987, 2014.

\bibitem{montalban2013power}
R.~Montalban, J.~A. L{\'o}pez-Salcedo, G.~Seco-Granados, and A.~L. Swindlehurst, ``Power allocation approaches for combined positioning and communications ofdm systems,'' in \emph{2013 IEEE 14th Workshop on Signal Processing Advances in Wireless Communications (SPAWC)}.\hskip 1em plus 0.5em minus 0.4em\relax IEEE, 2013, pp. 694--698.

\bibitem{wei20225g}
Z.~Wei~et al., ``5g prs-based sensing: A sensing reference signal approach for joint sensing and communication system,'' \emph{IEEE Transactions on Vehicular Technology}, vol.~72, no.~3, pp. 3250--3263, 2022.

\bibitem{ahmed2023sensing}
A.~Ahmed~et al., ``Sensing-centric isac signal processing,'' in \emph{Integrated Sensing and Communications}.\hskip 1em plus 0.5em minus 0.4em\relax Springer, 2023, pp. 179--209.

\bibitem{barneto2019full}
C.~B. Barneto~et al., ``Full-duplex ofdm radar with lte and 5g nr waveforms: Challenges, solutions, and measurements,'' \emph{IEEE Transactions on Microwave Theory and Techniques}, vol.~67, no.~10, pp. 4042--4054, 2019.

\bibitem{wymeersch2022radio}
H.~Wymeersch and G.~Seco-Granados, ``Radio localization and sensing—part ii: State-of-the-art and challenges,'' \emph{IEEE Communications Letters}, vol.~26, no.~12, pp. 2821--2825, 2022.

\bibitem{tigrek2012ofdm}
R.~F. Tigrek, W.~J. De~Heij, and P.~Van~Genderen, ``Ofdm signals as the radar waveform to solve doppler ambiguity,'' \emph{IEEE Transactions on Aerospace and Electronic Systems}, vol.~48, no.~1, pp. 130--143, 2012.

\bibitem{cover1999elements}
T.~Cover, \emph{Elements of information theory}.\hskip 1em plus 0.5em minus 0.4em\relax John Wiley \& Sons, 1999.

\bibitem{braun2014ofdm}
K.~M. Braun, ``Ofdm radar algorithms in mobile communication networks,'' Ph.D. dissertation, Karlsruhe, Karlsruher Institut f{\"u}r Technologie (KIT), Diss., 2014, 2014.

\end{thebibliography}

\end{document}